\documentclass[%
 reprint,
 amsmath,amssymb,aps,
]{revtex4-2}

\usepackage{graphicx}
\usepackage{dcolumn}%
\usepackage{bm}
\usepackage{hyperref}

\begin{document}
\title{Zitterbewegung\ CPT Violation in a Neutral Kaons System }
\author{J-M Rax}
\email{jean-marcel.rax@universite-paris-saclay.fr}
\affiliation{Universit\'{e} de Paris-Saclay\\
IJCLab-Facult\'{e} des Sciences d'Orsay\\
91405 Orsay France}
\date{\today}

\begin{abstract}
The observed $CP$ violation in neutral kaons experiments is explained as
an interplay between two oscillations in the earth's Schwarzschild geometry:
(i) mixing associated with second order weak coupling and (ii) strange
quark's zitterbewegung. This violation is in fact a $CPT$ violation with $T$ 
conservation rather than a $T$ violation with $CPT$ conservation. 
The Hermitian evolution of a stable kaons system leads to the identification of
this $CPT$ violation. Then, the finite lifetime of the short-lived kaon
rotates the violation parameter such that it appears as a $T$ violation.
\end{abstract}

\maketitle

Sixty years ago J. Christenson, J. Cronin, V. Fitch and R. Turlay reported
the first measurement of $CP$ violation \cite{1,2,3}. Following decades 
of precision measurements this small $CP$ asymmetry is now incorporated in the Standard Model.

The Lee, Oehme and Yang (LOY) model \cite{4,5}, based on two coupled
Schr\"{o}dinger equations, provides the canonical framework to describe
the dynamics of neutral kaons. In this letter we
revisit the influence of the earth gravitational field on neutral kaons
dynamics  \cite{6,7,8,9} and extend the LOY model to account for this influence.

We demonstrate that earth induces a $CPT$ violation with a $T$ conservation,
fully consistent with the experimental results and usually interpreted as a $%
CP$ violation with $T$ violation and $CPT$ conservation. 

First, assuming that the long-lived kaon $K_{L}$ and the short-lived one $%
K_{S}$ are stable particles, we find that the overlap of $CP$ eigenstates $%
\left\langle K_{S}\right. \left| K_{L}\right\rangle $ is an imaginary number
($T$ conservation and $CPT$ violation).

Then, taking into account the finite lifetime of $K_{S}$, the observed $%
\left\langle K_{S}\right. \left| K_{L}\right\rangle $ is rotated and becomes
a real number (apparent $T$ violation) with the value given by the
experimental data. The measurement of $\left\langle K_{S}\right. \left|
K_{L}\right\rangle $ providing this real value favored, in the past, the
inclusion of this gravitational effect as a $CP$ and $T$ violations in the
LOY model, rather than $CPT$ violation with $T$ conservation.

The details of the demonstration, the consequences of the results, and some
perspectives, can be found in a longer companion paper  \cite{10}. The
origin of direct $CP$ violation finds also its interpretation with this new
mechanism and the derivation of the parameter $\varepsilon ^{\prime }$ is
presented in the longer companion paper \cite{10}, this letter is restricted to $%
\varepsilon $. We set $\hbar
=c=1$. Most of the references are not recent because the subject of earth
as the origin of $CP$ violation did not attract much interest during the
past decades.

We follow the way Dirac resolved the apparent contradiction between the
requirements of relativistic dynamics and quantum mechanics: ({\it i})
energy-momentum dispersion relations are quadratic forms, however ({\it ii})
wave equations are linear forms with respect to energy/time derivation in
order to define a positive probability density  \cite{11}.\ Dirac developed a
first order quantum evolution such that the iteration of the time derivation
yielded the relativistic quadratic dispersion relation, in doing so he
identified the $\gamma $ matrices algebra.

In this letter, we use the iteration of the LOY first order wave equations
to set up coupled Klein-Gordon equations revealing the principal
gravitational coupling to be considered in the first order wave equations.

The main results of this letter are: ({\it i}) the origin of indirect $CP$
violation in kaons systems is identified, ({\it ii}) the measured value of
the indirect $CP$ violation parameter is predicted and ({\it iii}) the right
status of the symmetry violation induced by curvature and
quarks zitterbewegung is restored: $CPT$ violation with $T$
conservation rather than $T$ violation with $CPT$ conservation. $T$
violation appears as a result of the $K_{L}$ decays described as an
irreversible process with the Weisskopf-Wigner approximation \cite{12}. 

The rest frame evolution of a neutral kaons system ($K^{0}$/$\overline{K}%
^{0} $) is described by two amplitudes defining the state $\left| \Psi
\right\rangle $ as a function of the particle proper time $\tau $, 
\begin{equation}
\left| \Psi \right\rangle =a\left( \tau \right) \left| K^{0}\right\rangle
+b\left( \tau \right) \left| \overline{K}^{0}\right\rangle \text{.}
\label{ampK}
\end{equation}
The two $CP$ and energy eigenstates, $\left| K_{1}\right\rangle $ and $%
\left| K_{2}\right\rangle $ are the normalized sum and difference
of $\left| K^{0}\right\rangle $ and $\left| \overline{K}^{0}\right\rangle $.
The Weisskopf-Wigner approximation  \cite{12} describes decay through a
nonhermitian Hamiltonian $H_{K}$ driving a first order proper time
evolution $H_{K}\left| \Psi \right\rangle $ $=$ $j\hbar \partial _{\tau
}\left| \Psi \right\rangle $ with 
\begin{eqnarray}
H_{K} &=&Mc^{2}\left( \left| K^{0}\right\rangle \left\langle K^{0}\right|
+\left| \overline{K}^{0}\right\rangle \left\langle \overline{K}^{0}\right|
\right)  \nonumber \\
&&-\delta Mc^{2}\left( \left| \overline{K}^{0}\right\rangle \left\langle
K^{0}\right| +\left| K^{0}\right\rangle \left\langle \overline{K}^{0}\right|
\right) \text{.}  \label{kmh1}
\end{eqnarray}
This Hamiltonian looks like relativistic because of the occurrence of $c$,
but in fact it provides a description which is not fully relativistic. The
matrix elements $M$ $=$ $m-j\hbar \Gamma /2c^{2}$ and $\delta M$ $=$ $\delta
m-j\hbar \delta \Gamma /2c^{2}$ can be found in  \cite{13} and in the companion
paper \cite{10}.

In the following we restrict $H_{K}$ to its Hermitian part ($\Gamma =-\delta
\Gamma =0$) to avoid unphysical {\it interferences} between ({\it i})
phenomenological Weisskopf-Wigner damping ($\Gamma $,$\delta \Gamma $) and 
({\it ii}) first principles from quantum mechanics and general relativity.
The interplay between Weisskopf-Wigner decay and first principles is
the source of a confusion between $CPT$ and $T$ violations. In a
flat spacetime the Hermitian evolution ($\Gamma =-\delta
\Gamma =0$) of (\ref{ampK}) is  
\begin{equation}
j\frac{\partial }{\partial \tau }\left[ 
\begin{array}{l}
a \\ 
b
\end{array}
\right] =\left[ 
\begin{array}{cc}
m & -\delta m \\ 
-\delta m & m
\end{array}
\right] \cdot \left[ 
\begin{array}{l}
a \\ 
b
\end{array}
\right] \text{.}  \label{final0}
\end{equation}

The relativistic energy $\left( E\right) $-momentum $(p)$ relation is: $%
E^{2}-p^{2}=m^{2}$. To set up a quantum description ($E\leftrightarrow
j\partial /\partial \tau $) we have to consider the particle rest frame ($%
p=0 $) relation $\left\langle K^{0}\right| E\left| K^{0}\right\rangle =\pm m$%
. The relations used in Eq. (\ref{final0}) are the rest frame
restrictions ($p=0$) of the operator $+\sqrt{p^{2}+m^{2}}$ which is
non-local in a flat spacetime and can not operate in a curved
spacetime. A local evolution $E^{2}=p^{2}+m^{2}$ is more appropriate to work
in a curved spacetime.

An other critics of Eq. (\ref{final0}) is that nothing indicates that $%
\overline{K}^{0}$ is the antiparticle of $K^{0}$. These are two
particles with the same mass. This drawback is cured with the Feynman
prescription: if two additional velocity matrix elements are to be
considered, we must allocate a positive and a negative sign 
in order to describe one (the particle) propagation forward in time and one
(the antiparticle) propagation backward.

The earth Schwarzschild radius is $%
R_{S}$ $=$ $2GM_{\oplus }/c^{2}$ $=$ $8.87\times 10^{-3}$ m where $M_{\oplus
}$ is the earth mass and $G$ the gravitational constant. The earth radius $%
R_{\oplus }$ is far larger than $R_{S}\sim 10^{-9}R_{\oplus }$. In a
Schwarzschild spacetime, with time $t$ and radial coordinate $r$, the energy 
$E$ of a particle, with proper time $\tau $ at a distance $r$ $\ $from the
origin, is no longer $\ E=mdt/d\tau $ and is given by:
\begin{equation}
E=m\left( 1-R_{S}/r\right) dt/d\tau \text{.}  \label{energy}
\end{equation}
The radial position $r$ is the instantaneous {\it barycentric} position of
the mass/energy associated with the down $d/\overline{d}$ and strange $s/%
\overline{s}$ quarks dynamics inside the kaons. These quarks perform an
unknown, small scale, high frequency motion with respect to an average
center of mass/energy. This average center of mass/energy of the kaon
follows a geodesic: a very slow vertical free fall
combined with a fast (almost) horizontal  inertial motion. Moving to the horizontal
inertial frame of the kaon, the radial position $r$ is decomposed as $r=$ $%
R_{\oplus }+X+x$ the sum of: ({\it i}) the earth radius $R_{\oplus }$, plus (%
{\it ii}) the average vertical displacement $X\left( \tau \right) $, with
respect to $R_{\oplus }$, which is the average classical slow free fall
motion of the kaon ($\tau $ is the proper time of the kaon), plus ({\it iii}%
) the fast, unknown, internal vertical motion $x$, with respect to $%
R_{\oplus }+X$, of the instantaneous {\it barycentric} position of the
mass/energy associated with the quarks zitterbewegung dynamics. The
ordering $X+x\ll R_{\oplus }$ allows to expand the energy (\ref{energy}) in
the kaon frame where $t$ $=$ $\tau $: 
\begin{equation}
E=m\left[ 1-\left( R_{S}/R_{_{\oplus }}\right) +\left( R_{S}/R_{\oplus
}^{2}\right) \left( X+x\right) \right] \text{.}
\end{equation}
The proper time $\tau $ is the kaon rest frame time along a geodesic and $%
X\left( \tau \right) $ is the vertical part of this geodesic, so it can not
operate on the kaon internal state: in the kaon rest frame $X$  
is an additional time dependant energy.
The case is different for the quarks vertical
motions $x$ around $R_{\oplus }+X$ \ because $\tau $ is not the proper time
of the quarks. The internal position operator $\widehat{x}$ operate in the
tensorial product space of $d$ and $s$ Dirac's spinors states. Kaons are
spin $0$ diquark. We note: $\left| K^{0}\right\rangle =$ $\left| d%
\overline{s}\right\rangle $ and $\left| \overline{K}^{0}\right\rangle
=\left| \overline{d}s\right\rangle $. In the kaons Hilbert space ($\left|
K^{0}\right\rangle ,\left| \overline{K}^{0}\right\rangle $) $\widehat{x}$
is given by: 
\begin{widetext}
\begin{equation}
\widehat{x}=\left\langle d\overline{s}\right| \widehat{x}\left| d\overline{s}%
\right\rangle \left| K^{0}\right\rangle \left\langle K^{0}\right|
+\left\langle \overline{d}s\right| \widehat{x}\left| \overline{d}%
s\right\rangle \left| \overline{K}^{0}\right\rangle \left\langle \overline{K}%
^{0}\right| +\left\langle \overline{d}s\right| \widehat{x}\left| d\overline{s%
}\right\rangle \left| \overline{K}^{0}\right\rangle \left\langle
K^{0}\right| +\left\langle d\overline{s}\right| \widehat{x}\left| \overline{d%
}s\right\rangle \left| K^{0}\right\rangle \left\langle \overline{K}%
^{0}\right| \text{.}  \label{xxxx1}
\end{equation}
\end{widetext}
The dimension of the tensorial product of diquark spinor space $\left| q%
\overline{q^{\prime }}\right\rangle $ is $16$. The mass of the $s/\overline{s%
}$ strange component account for 96\% of the sum of the $d/\overline{d}$ and 
$s/\overline{s}$ components, thus we can neglect the dynamics of the $d/%
\overline{d}$ quark. The $s$ and $\overline{s}$ carry internal kinetic
and potential energy in addition to mass energy and this internal
energy is attached to the strange quark motion, the lighter quarks $%
d$ and $\overline{d}$ are just followers of the kaons slow free fall
dynamics and passive witness of the $s/\overline{s}$ fast zitterbewegung
motion. The Schr\"{o}dinger LOY equation on earth with $R_{S}\neq 0$ and $%
\Gamma =-\delta \Gamma =0$ is: 
\begin{widetext}
\begin{equation}
j\frac{\partial \left| \Psi \right\rangle }{\partial \tau }=m\left| \Psi
\right\rangle -\delta m\left[ \left| \overline{K}^{0}\right\rangle
\left\langle K^{0}\right| +\left| K^{0}\right\rangle \left\langle \overline{K%
}^{0}\right| \right] \left| \Psi \right\rangle -m\frac{R_{S}}{R_{\oplus }^{2}%
}\left[ R_{\oplus }-X\left( \tau \right) \right] \left| \Psi \right\rangle +m%
\frac{R_{S}}{R_{\oplus }^{2}}\ \widehat{x}\cdot \left| \Psi \right\rangle 
\text{.}  \label{scg2}
\end{equation}
\end{widetext}
The first and third terms on the right hand side of Eq. (\ref{scg2}) are
easily eliminated to obtain a more simple representation of the kaons state 
\begin{equation}
\left| \Phi \right\rangle =\left| \Psi \right\rangle \exp jm\left[ \tau -%
\frac{R_{S}}{R_{\oplus }}\left( \int_{0}^{\tau }du-\int_{0}^{\tau }\frac{%
X\left( u\right) }{R_{\oplus }}du\right) \right] \text{.}  \label{elim2}
\end{equation}
As $R_{S}/R_{\oplus }\sim 10^{-9}$ and $R_{S}X/R_{\oplus }^{2}\sim X\left[ 
\text{m}\right] \times 10^{-16}$ both gravitational terms in (\ref{elim2})
are negligible compared to $m\tau $ and unobservable with interferometric
experiments (they add up in the same way to the mass of $K_{S}$ and $K_{L}$%
). We can safely neglect the $R_{S}/R_{\oplus }$ term in (\ref{elim2}) and
define the amplitudes $u$ and $v$ of $\Phi $, 
\begin{equation}
\left| \Phi \right\rangle =\left| \Psi \right\rangle \exp jm\tau =u\left(
\tau \right) \left| K^{0}\right\rangle +v\left( \tau \right) \left| 
\overline{K}^{0}\right\rangle \text{.}  \label{cov23}
\end{equation}
Under the assumption of the strange quark dynamical dominance $u$ and $v$
fullfil 
\begin{eqnarray}
j\frac{\partial }{\partial \tau }\left[ 
\begin{array}{l}
u \\ 
v
\end{array}
\right] &=&\left[ 
\begin{array}{cc}
0 & -\delta m \\ 
-\delta m & 0
\end{array}
\right] \cdot \left[ 
\begin{array}{l}
u \\ 
v
\end{array}
\right] +m\frac{R_{S}}{R_{\oplus }^{2}}  \nonumber \\
&&\times \left[ 
\begin{array}{cc}
\left\langle \overline{s}\right| \widehat{x}\left| \overline{s}\right\rangle
& \left\langle \overline{s}\right| \widehat{x}\left| s\right\rangle \\ 
\left\langle s\right| \widehat{x}\left| \overline{s}\right\rangle & 
\left\langle s\right| \widehat{x}\left| s\right\rangle
\end{array}
\right] \cdot \left[ 
\begin{array}{l}
u \\ 
v
\end{array}
\right] \text{.}  \label{zut4}
\end{eqnarray}
where we have used Eqs. (\ref{xxxx1},\ref{scg2},\ref{elim2},\ref{cov23}).
The Compton wavelength ($\lambda _{C}=\hbar /mc=3.9\times 10^{-16}$ m) of
the kaon provides an upper bound of $\left\langle {}\right| \widehat{x}%
\left| {}\right\rangle $ in (\ref{zut4}) as quarks are bound states inside
the volume of the kaon. The value $R_{S}\lambda _{C}/R_{\oplus }^{2}$ $\sim $
$8.6\times 10^{-32}$ leads to the occurrence of a very strong ordering
fulfilled by these four matrix elements in front of $\delta m/m\sim
3.5\times 10^{-15}$.

Following Dirac's historical method, the fully relativistic dynamics
associated with (\ref{zut4}) is explored through an iteration of the
time derivation. Among the various gravitational terms we keep only the
dominant one, the others ($m^{2}R_{S}^{2}\lambda _{C}^{2}/R_{\oplus }^{4}$ $%
\delta m^{2}\sim 10^{-34}$ and $mR_{S}\lambda _{C}/\delta mR_{\oplus
}^{2}\sim 10^{-17}$) are negligible,

\begin{eqnarray}
-\frac{\partial ^{2}}{\partial \tau ^{2}}\left[ 
\begin{array}{l}
u \\ 
v
\end{array}
\right] &=&\left[ 
\begin{array}{cc}
\delta m^{2} & 0 \\ 
0 & \delta m^{2}
\end{array}
\right] \cdot \left[ 
\begin{array}{l}
u \\ 
v
\end{array}
\right] +jm\frac{R_{S}}{R_{\oplus }^{2}}  \nonumber \\
&&\times \left[ 
\begin{array}{cc}
\frac{\partial }{\partial \tau }\left\langle \overline{s}\right| \widehat{x}%
\left| \overline{s}\right\rangle & \frac{\partial }{\partial \tau }%
\left\langle \overline{s}\right| \widehat{x}\left| s\right\rangle \\ 
\frac{\partial }{\partial \tau }\left\langle s\right| \widehat{x}\left| 
\overline{s}\right\rangle & \frac{\partial }{\partial \tau }\left\langle
s\right| \widehat{x}\left| s\right\rangle
\end{array}
\right] \cdot \left[ 
\begin{array}{l}
u \\ 
v
\end{array}
\right] \label{eq12}
\end{eqnarray}

A straightforward calculation with Dirac's equation describing $s/\overline{s%
}$ quarks bounded in a kaon (presented and developed in the longer companion
paper  \cite{10}) demonstrates that 
\begin{equation}
\frac{\partial }{\partial \tau }\left[ 
\begin{array}{cc}
\left\langle \overline{s}\right| \widehat{x}\left| \overline{s}\right\rangle
& \left\langle \overline{s}\right| \widehat{x}\left| s\right\rangle \\ 
\left\langle s\right| \widehat{x}\left| \overline{s}\right\rangle & 
\left\langle s\right| \widehat{x}\left| s\right\rangle
\end{array}
\right] =\left[ 
\begin{array}{cc}
0 & c \\ 
c & 0
\end{array}
\right] \text{.} \label{ccc}
\end{equation}
This result belongs to the set of effects called {\it zitterbewegung } \cite{11}:
the Dirac position operator $\widehat{x}$ operating on spin $1/2$ fermions
spinors displays unfamiliar and nonintuitive properties, the associated
velocity operator (involved in the left hand side matrix in (\ref{ccc})) has
two eigenvalues $\pm c$ associated with four eigenvectors describing pairs
of fermion/antifermions \cite{11}. These velocity operator properties 
give (\ref{ccc}). This {\it %
zitterbewegung } behavior is the origin of the value of the non diagonal
elements in (\ref{ccc}). The diagonal elements are zero because the average
quark velocity in a bound state is zero \cite{10}.

We define the parameter $\kappa $: 
\begin{equation}
\kappa =\frac{m\hbar }{\delta m^{2}c}\frac{R_{S}}{4R_{\oplus }^{2}}=\left( 
\frac{m}{2\delta m}\right) ^{2}\frac{R_{S}\lambda _{C}}{R_{\oplus }^{2}}%
=1.7\times 10^{-3}\text{,}  \label{kj}
\end{equation}
With the {\it zitterbewegung} property (\ref{ccc}), equation (\ref{eq12}) is
then reduced to a simple square with an accuracy $O\left[ \kappa ^{3}\right]
\sim 10^{-9}$%
\begin{widetext}
    \begin{equation}
-\frac{\partial ^{2}}{\partial \tau ^{2}}\left[ 
\begin{array}{l}
u \\ 
v
\end{array}
\right] =\left( \pm \left[ 
\begin{array}{cc}
0 & \delta m \\ 
\delta m & 0
\end{array}
\right] \pm \delta m\left[ 
\begin{array}{cc}
2j\kappa & 2\kappa ^{2} \\ 
2\kappa ^{2} & 2j\kappa
\end{array}
\right] \right) ^{2}\cdot \left[ 
\begin{array}{l}
u \\ 
v
\end{array}
\right] +\delta m^{2}\times O\left[ 
\begin{array}{cc}
\kappa ^{4} & \kappa ^{3} \\ 
\kappa ^{3} & \kappa ^{4}
\end{array}
\right] \cdot \left[ 
\begin{array}{l}
u \\ 
v
\end{array}
\right] \text{.}  \label{eqf3}
\end{equation}
\end{widetext}
Despite the fact that this rest frame Klein-Gordon evolution is associated
with a local and complete relativistic dispersion, the right framework to
describe a quantum evolution is a set of
first order equations. The solutions of the second-order equation (\ref{eqf3}%
) includes redundant solutions which do not satisfy (\ref{zut4}). Among the
first order equations leading to (\ref{eqf3}) we have to consider one whose
limit is (\ref{final0}) when $\kappa =0$.

From (\ref{eqf3}) we go back to $\left( u,v\right) $ first order evolutions
and then to the initial kaons amplitudes $\left( a,b\right) $ with the help
of the change of variables from (\ref{cov23}) to (\ref{ampK}),

\begin{eqnarray}
j\frac{\partial }{\partial \tau }\left[ 
\begin{array}{l}
a \\ 
b
\end{array}
\right] &=&\left[ 
\begin{array}{cc}
m & -\delta m \\ 
-\delta m & m
\end{array}
\right] \cdot \left[ 
\begin{array}{l}
a \\ 
b
\end{array}
\right]  \nonumber \\
&&+\delta m\left[ 
\begin{array}{cc}
2j\kappa & -2\kappa ^{2} \\ 
-2\kappa ^{2} & -2j\kappa
\end{array}
\right] \cdot \left[ 
\begin{array}{l}
a \\ 
b
\end{array}
\right] \text{.}  \label{finalx}
\end{eqnarray}
As $\kappa \sim \partial \left\langle \overline{s}\right| \widehat{x}\left|
s\right\rangle /\partial \tau $ is proportional to a velocity matrix
element, to distinguish the $K^{0}$ from the $\overline{K}^{0}$, we have
implemented Feynman's prescriptions in going from (\ref{eqf3}) to (\ref
{finalx}) so that $a$ describe the amplitude of a particle whose
antiparticle is described by $b$. Moreover this ensures that the
energy eigenvalues are real. These energy eigenvalues of (\ref{finalx}) are: 
$m\mp \delta m$ with an accuracy $\delta m\times 10^{-12}$.\ 

The eigenvectors ($O\left[ \kappa ^{2}\right] $ terms are neglected)
associated with the time evolutions $\exp -j\left( m\pm \delta m\right) \tau
/\hbar $ are 
\begin{equation}
\left| K_{L/S}^{\oplus }\right\rangle =\frac{1\pm j\kappa }{\sqrt{2}}\left|
K^{0}\right\rangle \mp \frac{1\mp j\kappa }{\sqrt{2}}\left| \overline{K}%
^{0}\right\rangle \text{,}  \label{kl2}
\end{equation}
($L$ and $S$ are respectively associated with the upper and lower sign). A
better accuracy, neglecting $O\left[ \kappa ^{3}\right] $, can be easily
achieved \cite{10} but it does not change the results below. The very first
consequence of (\ref{kl2}) is the identification of the origin of $CP$
violation and its characterization as a $CPT$ violation with $%
T$ conservation 
\begin{equation}
\left. \left\langle K_{S}^{\oplus }\right. \left| K_{L}^{\oplus
}\right\rangle \right| _{\Gamma =-\delta \Gamma =0}=2j\kappa \text{.}
\label{keke}
\end{equation}
The two stables kaons $\left| \widetilde{K}_{L}\right\rangle $ and $\left| 
\widetilde{K}_{S}\right\rangle $ solution of Eq. (\ref{finalx}) evolve in
time according to 
\begin{equation}
\left| \widetilde{K}_{L/S}\right\rangle =\exp -j\left( m\pm \delta m\right)
\tau \left| K_{L/S}^{\oplus }\right\rangle \text{.}  \label{ls}
\end{equation}
Then, to work within the framework of the experimental conditions, we take into
account the $K_{S}$ decays. We assume that $K_{L}$ is stable, and not
depleted by $K_{S}$ gravitational regeneration, and that $K_{S}$ is unstable. 
In the following a star
index $\left( _{*}\right) $ indicates the {\it dressing} resulting from the
opening of the $K_{S}$ decay channels of the {\it bare} stable state $\left| 
\widetilde{K}_{S}\right\rangle $. The {\it dressed} kaons energy
eigenstates, $\left| K_{L*}^{\oplus }\right\rangle $ and $\left|
K_{S*}^{\oplus }\right\rangle $, are defined through relations similar to (%
\ref{ls}) 
\begin{equation}
\left| \widetilde{K}_{L*/S*}\right\rangle =\exp -j\left( m\pm \delta
m\right) \tau \left| K_{L*/S*}^{\oplus }\right\rangle \text{.}  \label{A3}
\end{equation}
Within the framework of the Weisskopf-Wigner approximation the evolution of $%
\widetilde{K}_{S*}$ is described by: ({\it i}) a decay with a rate $\left(
\Gamma -\delta \Gamma \right) /2\approx \Gamma \approx -\delta \Gamma $ and (%
{\it ii}) a gravitational regeneration with a rate $\left| \widetilde{K}%
_{L}\right\rangle \partial \left\langle \widetilde{K}_{L}\right. \left| 
\widetilde{K}_{S}\right\rangle /\partial \tau $ and (\ref{ls}) gives

\begin{equation}
\frac{\partial }{\partial \tau }\left\langle \widetilde{K}_{L}\right. \left| 
\widetilde{K}_{S}\right\rangle =2j\delta m\left\langle K_{L}^{\oplus
}\right. \left| K_{S}^{\oplus }\right\rangle \exp 2j\delta m\tau \text{.}
\label{rate2}
\end{equation}
The Schr\"{o}dinger equations for the dressed $K_{S*}$ is a balance between
decay and gravitational regeneration from a steady state $K_{L*}$ amplitude, 
\begin{eqnarray}
\frac{\partial \left| \widetilde{K}_{S*}\right\rangle }{\partial \tau }
&=&-j\left( m-\delta m\right) \left| \widetilde{K}_{S*}\right\rangle -\Gamma
\left| \widetilde{K}_{S*}\right\rangle  \nonumber \\
&&+\left| \widetilde{K}_{L*}\right\rangle \left( \frac{\partial }{\partial
\tau }\left\langle \widetilde{K}_{L}\right. \left| \widetilde{K}%
_{S}\right\rangle \right) \text{.}  \label{over2}
\end{eqnarray}
The observed experimental $CP$ violation parameter $\left\langle
K_{L*}^{\oplus }\right. \left| K_{S*}^{\oplus }\right\rangle $ is obtained
as the balance between the last two terms on the right hand side of (\ref
{over2}) 
\begin{equation}
-\Gamma \left\langle K_{L*}^{\oplus }\right. \left| K_{S*}^{\oplus
}\right\rangle +2j\delta m\left\langle K_{L}^{\oplus }\right. \left|
K_{S}^{\oplus }\right\rangle =0\text{,}
\end{equation}
where we have used (\ref{rate2}), (\ref{A3}) and $\left\langle
K_{L*}^{\oplus }\right. \left| K_{L*}^{\oplus }\right\rangle $ $=$ $1$.
Despite the fact that the {\it bare} parameter $\left\langle K_{L}^{\oplus
}\right. \left| K_{S}^{\oplus }\right\rangle $ is an imaginary number, the
measured $CP$ violation parameter $\left\langle K_{L*}^{\oplus }\right.
\left| K_{S*}^{\oplus }\right\rangle $ is a real number because of $K_{S}$
finite lifetime, 
\begin{equation}
\left. \frac{\left\langle K_{L*}^{\oplus }\right. \left| K_{S*}^{\oplus
}\right\rangle }{2}\right| _{\Gamma = -\delta \Gamma \neq 0}=\kappa 
\frac{2\delta mc^{2}}{\hbar \Gamma }=1.6\times 10^{-3}\text{,}  \label{ep}
\end{equation}
in good agreement with the experimental $%
\mathop{\rm Re}%
\left( \varepsilon \right) $  \cite{14}. Note that, if the restriction of the
dynamics to the strange quark is taken into account through a lowering of
the mass $m$ to $\left( m_{s}/m_{d}+m_{s}\right) \times m$, this gives $%
1.60\times 10^{-3}$.

The numerical coincidence $mg\hbar /\left( 2\delta m\right) ^{2}c^{3}\approx 
\mathop{\rm Re}%
\varepsilon /2$ 
($g=c^{2}R_{S}/2R_{\oplus }^{2}=9.8$ m/s$^{2}$) was
identified four decades ago by E. Fischbach  \cite{15}.
Fischbach was exploring the impact of gravitation and he noted the puzzling
relation: a simple {\it ad hoc} increase by a factor $m/2\delta m$ of the
coupling $\hbar g/2\delta mc^{3}$ gives $%
\mathop{\rm Re}%
\varepsilon /2$.

At the fundamental level the $CP$ violation observed on earth in the $%
K_{L}/K_{S}$ system is a $CPT$ violation and the $K_{S}$ finite lifetime
makes it appear as a $T$ violation.

Finally the $K_{L}$ instability can also be taken into account with the
Bell-Steinberger unitarity relations  \cite{16}. The result is $\arg \left(
\varepsilon \right) \approx 43,4^{\circ }$  \cite{10} in good agreement with the
experimental data  \cite{14}. The origin of direct violation finds also an
explanation within the framework of this new mechanism. We report
here the results obtained in the companion paper  \cite{10}: if the parameters are
expressed in a phase convention independent format then the ratio $%
\mathop{\rm Re}%
\left( \varepsilon ^{\prime }/\varepsilon \right) $ is $1.6\times 10^{-3}$.

The simplicity of the mechanism and the agreement between the predicted
values of the $CP$ violation parameters and the experimental data allow to
conclude that the set of relations Eqs. (\ref{finalx},\ref{kl2},\ref{over2},%
\ref{ep}) is an operational framework to explain $CP$ violations in kaons
experiments \ on earth, provided that: ({\it i}) the dissipative dressing of
the various amplitude and ({\it ii}) their dependance/independence on
initial phase conventions are carefully taken into account.

This new mechanism of $CP$ violation results from the interplay between two
oscillations: {\it (i) }quarks's zitterbewegung inherent to
fermion/antifermion pairs and ({\it ii}) kaons mixing associated with second
order weak interaction coupling. As the frequencies of these two
oscillations are rather different, this is a secular effects like the
ponderomotive shift/force  \cite{17} which is usually described in term of an {\it %
effective} or a {\it dressed} mass in analytical mechanics  \cite{18}.

This mechanism of {\it zitterbewegung\ CP violation} might change the way we
understand how our matter-dominated universe emerged during its early
evolution. The observed present level of $CP$ violation on earth is at least
ten orders of magnitudes smaller than the one needed to explain the observed
matter-antimatter asymmetry. At a distance $R$ from the center of a massive
object, or a large energy fluctuation, with Schwarzschild radius $R_{S}$,
the amplitude of the $CP$ violation parameter (\ref{ep}) is the product of
three factors: $\left( m/2\delta m\right) ^{2}$ and $\left( 2\delta
mc^{2}/\hbar \Gamma \right) $ which are associated with the fundamental
interactions (electromagnetic, weak and strong), and $\left( R_{S}\lambda
_{C}/R^{2}\right) $ which is associated with spacetime and can be far larger
than its earth value. Larger perspectives are presented in the longer
companion paper  \cite{10}. 

\begin{acknowledgments}
    This work was undertaken during a sabbatical year at Princeton and completed
    at the University of Paris-Saclay. The financial support of a G. R.
    Andlinger fellowship from Princeton University ACEE is acknowledged. The
    author gratefully thanks Prof. N. Fisch and his team, Drs. E.J. Kolmes, M.E.
    Mlodik, I.E. Ochs and T. Rubin, for their warm welcome in their aneutronic
    fusion project and for the intellectual atmosphere which provided him the
    ideal conditions for simultaneously addressing the issues of \ aneutronic
    fusion and this long-standing $CP$ problem.
\end{acknowledgments}

\end{document}